\documentstyle[12pt,aaspp4]{article}

\def\etal{{et al.}}
\def\asca{{\it ASCA}}

\def  \Mo          {$ M_{\odot} $}     
\newbox\grsign \setbox\grsign=\hbox{$>$} 
\newdimen\grdimen \grdimen=\ht\grsign
\newbox\simlessbox \newbox\simgreatbox \newbox\simpropbox
\setbox\simgreatbox=\hbox{\raise.5ex\hbox{$>$}\llap
     {\lower.5ex\hbox{$\sim$}}}\ht1=\grdimen\dp1=0pt
\setbox\simlessbox=\hbox{\raise.5ex\hbox{$<$}\llap
     {\lower.5ex\hbox{$\sim$}}}\ht2=\grdimen\dp2=0pt
\setbox\simpropbox=\hbox{\raise.5ex\hbox{$\propto$}\llap
     {\lower.5ex\hbox{$\sim$}}}\ht2=\grdimen\dp2=0pt
\def\simgreat{\mathrel{\copy\simgreatbox}}

\begin{document}

\title{X-ray Observations of Mrk 231} 

\author {T.J.Turner\altaffilmark{1,2,3} \\
 } 

\altaffiltext{1}{Laboratory for High Energy Astrophysics, Code 660,
	NASA/Goddard Space Flight Center,
  	Greenbelt, MD 20771}
\altaffiltext{2}{Universities Space Research Association}
\altaffiltext{3}{Present address, University of Maryland, Baltimore County}
 
\slugcomment{To be submitted to {\em The Astrophysical Journal}}

\begin{abstract}
This paper presents new X-ray observations of Mrk~231, an active galaxy  
of particular interest due to its large infrared luminosity and 
the presence of several 
blueshifted broad absorption line (BAL) systems, a phenomenon observed in 
a small fraction of QSOs.  

A {\it ROSAT} HRI image of Mrk~231 is presented, this shows an 
extended region of soft X-ray emission, 
covering several tens of kpc, consistent with 
the extent of the host galaxy.  
An {\it ASCA} observation of Mrk~231 is also presented. 
Hard X-rays are detected but the data show no
significant variability in X-ray flux. 
The hard X-ray continuum is  heavily attenuated 
and X-ray column estimates  range from 
$ \sim 2 \times 10^{22} -  10^{23} {\rm cm}^{-2}$ depending on whether 
the material is assumed to be neutral or ionized, and on 
the model assumed for the extended X-ray component. These {\it ASCA} data 
provide only the second hard X-ray spectrum of a BAL AGN presented to date. 
The broad-band spectral-energy-distribution of the source is discussed. While 
Mrk~231 is X-ray weak compared to Seyfert~1 galaxies, it has an 
optical-to-X-ray spectrum typical of a QSO. 

\end{abstract}


\section{Introduction}
\label{sec:intro} 

Recent observations of luminous infrared  galaxies (LIGs)
indicate that most occur in interacting systems. It is thought that 
galaxy interactions can fuel Seyfert-type activity in galactic nuclei 
(e.g. Xia \etal\ 1998 and references therein) and also result 
in prodigious starburst activity, 
often observed close to the nucleus.  
Observations in the infrared (IR) band show strong emission from 
galaxies dominated by starburst activity. 
It is often difficult to determine whether these galaxies also
have active nuclei, based on the IR data alone. 
Radio and X-ray observations help clarify the picture. 
The amount of radio and X-ray emission expected from 
the observed starburst regions can be predicted, from 
observation of the phenomenon in normal galaxies (David \etal\ 1992). 
Some authors suggest that LIGs represent an early phase of evolution for 
active galactic nuclei  (AGN); 
Lipari, Terlevich and Macchetto (1993) found that most 
AGN which have strong Fe{\sc ii} emission are also LIGs. Mrk~231 
(z=0.042) is 
an ultraluminous infrared galaxy, 
and one of the strongest known Fe{\sc ii} emitters. 
Furthermore, Mrk~231 has one of the highest bolometric luminosities 
in the local 
(z $<$ 0.1) universe (Sanders \etal\ 1988) with $M_v=-22.5$ 
(Rieke \& Low 1972), 
$L_{8-1000 \mu m} \sim 4 \times 10^{12} L_{\odot}$ (Soifer \etal\ 1986),  
$L_{IR}/L_B \sim 200$ (Lipari, Colina \& Macchetto  1994) and 
$L_{BOL}> 10^{46} {\rm erg\ s}^{-1}$ (Soifer \etal\ 1987).  
Boksenberg \etal\ (1977) show that an active nucleus exists in 
Mrk~231; broad emission lines are observed from permitted 
transitions, the species and line widths are similar to those 
defining the Seyfert~1 class. However, 
no narrow emission lines 
are seen except for O\verb+[+{\sc ii}\verb+]+ $\lambda 3727$. 
This O\verb+[+{\sc ii}\verb+]+ emission line is probably produced in gas which is 
ionized by hot young stars (Smith \etal\ 1995) 
rather than by ultraviolet radiation from the nucleus, since 
the latter is unlikely to 
emerge through the dense circumnuclear absorber. 
The absence of narrow emission lines is sometimes observed in quasi-stellar objects (QSOs)  
but not in Seyfert galaxies. Another extraordinary property 
of Mrk~231 is the 
high polarization, $\sim 20\%$ at 2800 \AA\ (e.g. Smith 
\etal\ 1995, after accounting for dilution from hot stars). 
In this case the 
polarized flux is thought to be nuclear emission scattered from 
circumnuclear dust (Smith \etal\ 1995). 

Mrk~231 is radio-quiet but 
Smith, Lonsdale \& Lonsdale (1998) find the strength and compactness of the 
VLBI (milli-arcsecond) radio emission to infer the presence of an 
 active nucleus.  
Bryant \& Scoville  (1996) find that starburst activity can 
only account for 40\% of the 
IR luminosity in Mrk 231, supporting the evidence for an active nucleus, 
which must power the remaining emission.  {\it HST} data 
showed a compact nucleus enclosed 
on the south side by a dense arc of star-forming knots concentrated 
$\sim 3 - 4''$ away (Surace \etal\ 1998). 

It has also been suggested that there is a connection between 
the presence of strong Fe{\sc ii} emission and blueshifted absorption 
systems (Hartig \& Baldwin 1986;  Weymann \etal\ 1991; Lawrence \etal\ 1997), 
although the physical origin of this connection is unclear. 
Mrk~231 has several blueshifted 
Broad Absorption Line (BAL) systems evident in the ultraviolet data. 
Adams (1972), Boksenberg \etal\ (1977), Arakelian \etal\ (1971), Boroson \etal\ (1991), 
Rudy \etal\ (1985)  and others 
discuss three distinguishable absorption systems in Mrk~231. The most 
prominent is defined by Na{\sc i}, Ca{\sc ii} and He{\sc i} lines 
originating in gas with an outflow velocity $\sim 4200\ {\rm km\ s}^{-1}$ 
relative to the line emission seen in the source. 
A second system shows only the D lines of Na{\sc i} and an 
outflow velocity $\sim 6000\ {\rm km\ s}^{-1}$. A third system 
is observed in Balmer and Ca{\sc ii}  absorption lines and has a 
redshift implying inflow at a velocity $\sim 200\ {\rm km\ s}^{-1}$.
A fourth system appeared between 1984 December and 1988 May, 
with prominent Na{\sc i} D and He{\sc i} absorption at 
an outflow velocity of $\sim 8000\ {\rm km\ s}^{-1}$  (Boroson 
\etal\ 1991, Kollatschney, Dietrich \& Hagan 1992). Further 
absorption variability was observed when a Na{\sc i} D line 
disappeared between 1991 and 1994 and the profile of some lines 
varied (Forster \etal\ 1995). 

Rudy \etal\ (1985) showed that the shapes of absorption features and velocities in the 
absorbing gas systems of Mrk~231 are the same as those 
observed in BAL QSOs and 
the gas is probably accelerated by the same mechanism. Thus 
Mrk~231 offers an opportunity to study the BAL phenomenon at 
low redshift, and with a component containing species of unusually low 
ionization-state. Most BAL systems are dominated by 
high-ionization species such as C{\sc iv}$\lambda \lambda 1548$,1551; 
Si{\sc iv} $\lambda \lambda 1394$,1403 and N{\sc v}$\lambda \lambda 1239$, 1243 
and only a small percentage of sources ($\sim 15\%$) 
show absorption by low-ionization species, such as 
Mg{\sc ii} $\lambda \lambda 2796$, 2803; Al{\sc ii}$\lambda 1671$; 
Al{\sc iii}$\lambda \lambda 1855$, 1863 and C{\sc ii}$\lambda 1335$ 
(Voit, Weymann \& Korista 1993). Absorption by 
Na {\sc i} D, He {\sc i}$\lambda 3889$ and Mg {\sc i}$\lambda 2853$ 
as observed in Mrk~231, is quite rare. 
The detection of a 10$\mu m$ silicate absorption feature indicates that dust 
is an important constituent of at least one of the absorbing systems in 
Mrk~231 (Roche, Aitken \& Whitmore 1983). 
Forster \etal\ (1995) conclude the absorption systems are most likely 
associated with Mrk~231, rather than intervening galaxies, although they 
find no evidence these systems are linked to a starburst-superwind in this
case. Rudy \etal\ (1985) also point out that the axial ratio in this source 
is 0.48 (minor/major), and so the source is observed too close to a face-on 
orientation for the BAL gas to be consistent with interstellar 
clouds in the plane of the host galaxy. 
Mrk~231 also contains an OH maser (Baan 1985) and a CO emission region to
the nucleus  (Scoville \etal\ 1989). Bryant \& Scoville (1996) estimate that
$\sim 3.4 \times 10^9$ \Mo of gas lies within a radius 420 pc of the
nucleus and the material is most likely in the form of a disk.

New X-ray observations of Mrk~231 are presented in this paper. In \S 2 the
{\it ROSAT} HRI image of the source is shown, and the soft X-ray spectrum
observed by the PSPC is discussed.  In \S 3 an {\it ASCA} observation of
Mrk~231 is presented, which provides the first hard X-ray observation of this 
source, and only the second hard X-ray spectrum of a BAL
AGN in the literature.

\section{The ROSAT Observations}

Mrk~231 was observed by the {\it ROSAT} High Resolution Instrument (HRI) 
between 1996 May 12 and 1996 June 11, with a total 
exposure time of 31 ks. Fig~1 shows the HRI image, 
in $1''$ bins, smoothed with a Gaussian of width $\sigma=1.5''$. 
A white cross shows the position of the optical and radio nucleus 
(Surace \etal\ 1998) and the white bar shows an extent of $10''$.  
Two X-ray peaks are evident, spaced $\sim 13''$ apart. There is also
the suggestion of asymmetric extent, with an extended region of X-ray flux 
west of the peaks. 
There are several other bright point sources in the HRI image, 
and two were identified as the AGN 
\verb+F219_015+ and \verb+F219_045+. The optical 
positions (obtained from NED, which contains a p.comm.   
from Keith Mason) and the HRI positions for these AGN agreed to within 
$\sim 3''$, indicating an accurate aspect-solution for this image. 
The optical position of the nucleus 
is coincident with the region of relatively low X-ray flux, 
rather than the X-ray peaks in Fig.~1. There are several other serendipitous
point sources in the HRI image and none show the X-ray extent evident in
Mrk~231 demonstrating that the extended emission observed
for Mrk~231 is real and not an artifact of attitude reconstruction problems in
the HRI image.

The {\it HST} image of Mrk~231 shows a great deal of structure 
(Surace \etal\ 1998, c.f. their Fig.~7), most notably 
a region of emission to the NW of the Seyfert nucleus, composed of 
knots of emission from star formation. Another distinct peak 
in the {\it HST} data lies $\sim 4''$ south of the 
nucleus, again due to star formation. The extended emission in the 
X-ray image seems to be similar in shape, but on a size-scale 
a factor of $\sim 3$ larger than that of the bright 
star-forming knots evident in the {\it HST} image. 
At a redshift z=0.042, $1''$ corresponds to 
$\sim$ 1 kpc 
(assuming H$_{\rm 0}=50 {\rm km s^{-1} Mpc^{-1}}$, q$_{\rm 0}=0.5$) 
and so the HRI image implies emission across a total diameter of 
$\sim 30$ kpc. Comparison with the {\it HST} image of Surace 
\etal\ (1998; their Fig.~1) shows the X-ray emission matches the 
optical extent of the host galaxy. 
The total flux was $F(0.5-2\ {\rm keV})=1^+_-0.1 \times 10^{-13} {\rm erg\ cm}^{-2} 
{\rm s}^{-1}$ 
(assuming the spectral model derived from the PSPC fit, as noted below). 
The luminosity in the 
observed X-ray emission is $L(0.1-2\ {\rm keV}) \sim 10^{42} {\rm erg\ s}^{-1}$ 
and so is higher than expected from the summed emission of 
stars and hot gas in a normal galaxy, but may be related to the 
starburst activity. The nucleus appears relatively weak in the 
0.1 -- 2 keV band, perhaps because it is more heavily 
absorbed than emission from regions further out in the galaxy. 

Light curves were extracted from circular regions of radius $6''$ 
centered on each of the X-ray peaks. No significant variability was 
evident in the time series from either region. The flux of the 
two peaks is almost equal, in the soft X-ray band. The flux of the 
NW peak is $47^+_-9$ \% of the total flux in the pair. 

The {\it ROSAT} PSPC observed Mrk~231 1991 June 7 -- 8 for a total
of 24 ks. The two peaks evident in the HRI image cannot be resolved 
using the PSPC data (which has a point-spread-function of 
width $FWHM \sim 25''$). 
The data were corrected for time-dependent effects using the tool 
{\sc pcpicor v 2.2.0}.  Source data were extracted 
from a cell of radius $1.3'$ and  background data from an 
annular region centered on the source (but excluding some parts of 
the image contaminated by serendipitous sources).  
The PSPC spectrum has already been presented by Rigopoulou, Lawrence and 
Rowan-Robinson (1996). Those authors noted a poor fit to a power-law model,  
attenuated by a column of neutral material. That result is confirmed here, 
yielding $\chi^2=42$ for 18 degrees of freedom ($dof$). 

The fit is significantly improved (at $>99\% $ confidence) 
by addition of a narrow 
gaussian line with rest-energy $E=0.86^+_-0.06$ keV and 
equivalent width $EW=490^+_-190$ eV 
($\chi^2/dof=20.0/15$)  confirming the Rigopoulou \etal\ (1996) 
result and 
yielding a flux $F(0.5-2\ {\rm keV})=1 \times 10^{-13} {\rm erg\ cm}^{-2}
{\rm s}^{-1}$. This flux is in excellent agreement with that from the same 
region in the HRI observation, as expected given the extent of the  
emission. Strong line emission would be expected if the soft X-rays 
are from a region of starburst emission, and, as the image indicates the 
soft band is dominated by the emission from an extended region, 
the data were fit with a Raymond \& Smith model (1977); this model 
describes the spectrum of radiation emitted by a hot optically 
thin plasma with abundances and equilibrium ionization
balance appropriate to interstellar conditions. 
A single-temperature 
plasma is, again, a poor fit to the data ($\chi^2/dof=42.7/18$). 
The possibility of some contribution from the 
nuclear component was considered. Adding a power-law to the fit 
yielded a photon index $\Gamma=1.73^{+1.14}_{-0.27}$ and 
$\chi^2/dof=19.6/20$.  The absorption to both 
spectral components was found to be 
consistent with the Galactic line-of-sight value 
$N_H=1.03 \times 10^{20} {\rm cm}^{-2}$ (Elvis, Lockman \& Wilkes 1989). 
Such a low absorption is consistent with the origin of the
soft X-ray emission from regions outside of the nucleus. The 
nuclear component could be observed as scattered nuclear light, or could 
represent an unattenuated fraction of the nuclear continuum 
viewed through holes in the absorber. 
The parameterization using a Raymond-Smith plasma plus power-law 
is denoted ''Case-1''. 

Alternatively, the single-temperature 
plasma may be an inadequate model for the starburst region. 
A two-temperature solution yielded a significant improvement to the fit 
with $\chi^2/dof=22.6/16$,  a low temperature component 
$kT=0.31^{+0.40}_{-0.03}$keV, with normalization 
$n=2.3^{+1.3}_{-0.4} \times 10^{-5}$ and hot component 
$kT=1.8^{+60}_{-1.8}$keV, 
$n=8.3^{+13.2}_{-4.0} \times 10^{-5}$. The normalization is in units 
 $10^{-14}/4\pi D^2\int n_en_hdV$
where $D$ is the luminosity distance to the source (cm), $n_e$ is the
electron density (cm$^{-3}$) and $n_h$ is the hydrogen density (cm$^{-3}$). 
The parameterization using a two-temperature 
Raymond-Smith plasma is denoted ''Case-2''.

\section{The ASCA Observations} 

The {\it ASCA} solid-state imaging spectrometers (SISs) and 
gas imaging spectrometers (GISs) cover the 
$\sim$ 0.4 -- 10 keV and  $\sim$0.8 -- 10~keV bandpasses, respectively. 
Mrk 231 was observed by {\it ASCA} 1994 Dec 5. 
The data were reduced in the same way as the Seyfert galaxies
presented in Nandra \etal\ (1997) and Turner \etal\ (1997). For details of 
the data reduction method and instruments see Nandra \etal\ (1997). Data screening 
yielded effective exposure times of $\sim 20 $ ks in all four instruments.  
Examination of the {\it ASCA} images reveals marginal evidence for 
\verb+F219_045+ in the GIS image, but no evidence for hard X-ray emission 
from any of the other sources seen in the {\it ROSAT} images. 
\verb+F219_015+ and \verb+F219_045+ 
are not close enough to be confused with Mrk~231, but there are 
other sources of contamination as discussed below. 

{\it ASCA} light curves from each instrument were examined over the 0.5-10 keV
range, using time bins covering 128s to 5760 s. Mrk~231 was weak and the
background light curve showed some low-amplitude variations, however,  no flux
variations were evident within the {\it ASCA} observation, which could be
attributed to the AGN.

The source flux was 
$F \sim (1.0^+_-0.3) \times 10^{-12} {\rm erg\ cm}^{-2} {\rm s}^{-1}$ 
in the 2 -- 10 keV band corresponding to an observed luminosity 
$L(2-10\ {\rm keV})=8 \times 10^{42} {\rm erg\ s}^{-1}$. The soft flux is 
$F \sim (1.9^+_-0.5) \times 10^{-13} {\rm erg\ cm}^{-2} {\rm s}^{-1}$ 
in the 0.5 -- 2 keV band. This soft flux appears higher than 
that observed by {\it ROSAT}, but the $psf$ of the {\it ASCA} 
instruments is significantly wider than that of the {\it ROSAT} 
instruments. This is results in some contamination of the soft part of
the {\it ASCA} spectrum by unidentified sources close to Mrk~231, evident in
the PSPC and HRI images, as discussed in the following section.

\subsection{The ASCA Spectra} 

{\it ASCA} SIS data below an energy of 0.60 keV (0.66 keV rest-frame) were
excluded from the spectral analysis as it is commonly accepted that there are
uncertainties associated with the calibration in that band.  The {\it ASCA}
spectra were extracted using circular regions of radii $3.2'$ and $6.8'$ for
the SIS and GIS instruments, respectively. The PSPC image was then used to
examine the sources of contamination in each case. The SIS extraction cell
encompasses one source of contamination, and the GIS cell encompasses three.
These sources were faint and unidentified, so it is difficult to model the
precise contamination from each. However, 
it was found the contaminating flux from these serendipitous sources 
could be {\it parameterized} as 45\% (SIS) and 112\% (GIS) increases in the 
normalization of the thermal component ($kT=0.31$ keV) used to model 
the circumnuclear emission in Mrk~231.

\subsubsection{Case-1}

In Case-1 the soft X-ray flux was assumed to originate in 
a single-temperature Raymond-Smith plasma 
with some contribution from the nuclear continuum, viewed through unattenuated 
lines-of-sight or via scattering. 
The temperature and normalization of the Raymond-Smith plasma 
were fixed at the values 
determined from fitting the PSPC data, with 
an adjustment to account for the soft flux from contaminating sources, 
as noted above. 

The nuclear power-law was modeled assuming 
attenuation by a column of neutral material, $N_H$, 
at the redshift of the source and  
with a fraction of the emission assumed to be 
observed without attenuation. Again, 
an additional column was fixed at the Galactic line-of-sight value
$N_H=1.03 \times 10^{20} {\rm cm}^{-2}$ (at a redshift of zero).  
This fit yielded a photon index $\Gamma \sim 1.9$, with a column 
$N_H \sim 9 \times 10^{22} {\rm cm^{-2}}$ covering 79\% of the source 
(fit 1, Table~1). 

If the circumnuclear material is ionized, then the column could be 
higher than that indicated by fits which assume a neutral absorber. 
The data were fit using 
models based upon the photoionization code ION (Netzer 1996), 
once again, an unabsorbed thermal component of fixed 
temperature and normalization was allowed, the results are shown in Table~1 
(fit 2). 
If the column is assumed to be fully-covering the source, but ionized, then
the inferred index is $\Gamma \sim 1$. In its simplest interpretation this
represents an unusually flat index for the primary continuum.  This may  
indicate another component is present, 
such as that due to Compton reflection,  flattening 
the observed spectrum. It is not possible to obtain useful constraints on 
Compton reflection in Mrk~231 (using these data).  Therefore the 
fits were repeated with the photon index fixed at two values often found for
Seyfert galaxies (e.g. George \etal\ 1998 and references therein). 
This allowed an investigation of the nuclear attenuation 
under two simple assumptions, $\Gamma=1.5$ and $\Gamma=2$. 
In both these cases, the nuclear absorption was found 
to be $\sim 10^{23} {\rm cm^{-2}}$ (fits 3-4, Table~1). 

A small contribution to iron K$\alpha$ emission 
is expected from the thermal component, and 
is visible in the model plot (Fig~2, top panel).  However, 
no significant line emission was detected from the K-shell of iron, 
(although a high point on the data/model ratio is duely noted, Fig.~2, bottom panel). 
These data 
yield an upper limit (90\% confidence) on the equivalent width to be 
$EW < 800 $ eV. The {\it ASCA} data do not show a significant line 
at 0.86 keV, as indicated by the 
PSPC data, although the spectra are consistent with the presence of a 
line of the strength inferred by the PSPC data.

\subsubsection{Case-2}

In Case-2 the starburst region was assumed to be parameterized as a 
two-temperature Raymond-Smith plasma with the temperatures fixed 
at $kT=0.31$ and $kT=1.8$ keV, as found in \S 2. 
In this case a two-temperature component was allowed in the model, 
with normalizations fixed at the values derived from the PSPC data (again  
with an adjustment to the normalization of the cool component to 
account for flux from the contaminating sources). This 
parameterization left less soft 
 flux to be attributed to the nucleus and hence inferred different 
fit parameters to Case-1. 

A power-law was considered, attenuated by a column 
of neutral material at the redshift of the source,  again 
an additional column was fixed at the Galactic line-of-sight value. 
No unattenuated fraction was allowed. This fit yielded a photon index 
$\Gamma \sim 0.6$ (fit 5, Table~1) and the fits were repeated with 
$\Gamma=1.5$ and $\Gamma=2$ as before.  
By fixing 
the photon index a more meaningful estimate of the column may be obtained, 
$N_H \sim 2 - 4 \times 10^{22} {\rm cm}^{-2}$ (fits 6 -- 7). 
Fits assuming an ionized absorber yielded column estimates 
$N_H \sim 5 \times 10^{22} - 10^{23} {\rm cm}^{-2}$. 
Whether the absorber is neutral or ionized, Mrk~231 clearly has a hard 
X-ray continuum component, attenuated by a column consistent with that 
observed in many Seyfert~2 galaxies (Turner \etal\ 1997).

\begin{deluxetable}{llcccc}

\tablecaption{{\it ASCA} Spectral Fits}

\tablehead{
\colhead{Fit} &
\colhead{ $\Gamma$ } & \colhead{$A^a$} 
& \colhead{$N_H^b$}
&  \colhead{$U_X^c$/Frac} 
& \colhead{$\chi^2 /dof$} 
}
\startdata
\hline
\multicolumn{6}{c}{Case-1: Neutral Absorber} \nl
\hline
1& $1.89^{+1.09}_{-0.56}$  &  $5.20^{+30.50}_{-5.20}$ &  $9.1^{+98.9p}_{-9.10}$
	& $0.79^{+0.17p}_{-0.79p}$  & 94.0/98  \nl
\hline
\multicolumn{6}{c}{Case 1: Ionized Absorber} \nl
\hline
2& $1.10^{+0.78}_{-0.44}$  &  $1.10^{+2.67}_{-0.48}$ &  $15.1^{+24.9}_{-15.1}$ 
	& $8.91^{+1.09}_{-8.91}$  & 94.0/98  \nl 
3& $1.5(f)$  & $2.09^{+0.61}_{-0.68}$ &  $11.7^{+19.8}_{-10.8}$ 
	& $2.88^{+5.84}_{-2.88}$  & 99.2/99  \nl
4& $2.0(f)$  & $4.42^{+1.50}_{-1.50}$  &  $11.2^{+9.9}_{-8.0}$ 
	& $1.71^{+1.24}_{-1.20}$  & 104.8/99  \nl
\hline
\multicolumn{6}{c}{Case 2: Neutral Absorber} \nl
\hline
5& $0.64^{+0.97}_{-0.48}$  &  $0.45^{+1.87}_{-0.24}$ &  $0.2 < 2.7$
	& \nodata  & 96.3/99  \nl
6& $1.5(f)$  & $1.86^+_-0.45$ &  $2.1^{+1.7}_{-1.0}$ 
	&  \nodata  & 100/100  \nl
7& $2.0(f)$  & $4.39^{+1.57}_{-1.20}$  &  $3.7^{+2.7}_{-1.5}$ 
	&  \nodata  & 103.6/100  \\
\hline
\multicolumn{6}{c}{Case 2: Ionized Absorber} \nl
\hline
8& $1.03^{+1.50}_{-0.92}$  &  $0.92^{+9.65}_{-0.73}$ &  $5.5^{+29.9}_{-5.5}$ 
	& $1.22^{+8.78}_{-1.22}$  & 95.5/98  \nl
 9&$1.5(f)$  & $2.12^{+0.73}_{-0.68}$ &  $10.5^{+16.5}_{-8.6}$ 
	& $1.45^{+2.33}_{-1.45}$  & 96.1/99  \nl
 10&$2.0(f)$  & $4.69^{+2.05}_{-1.43}$  &  $12.6^{+15.6}_{-8.4}$ 
	& $1.35^{+1.45}_{-1.35}$  & 98.2/99  \nl
\hline

\tablenotetext{a} {The power-law normalization at 1 keV, in 
units $10^{-4}$ photons s$^{-1}$ keV$^{-1}$.} 
\tablenotetext{b} {Absorbing column in units $10^{22}\ {\rm cm^{-2} }$, at 
the redshift of the source. An additional column of 
$1.03 \times 10^{20} {\rm cm}^{-2}$ 
was included in the fit, at redshift z=0.} 
\tablenotetext{c} {Ionization parameter or covering fraction} 
\tablenotetext{} { {\it p} indicates an error estimation 
hit the limit allowed for that parameter.}
\tablenotetext{} {An unattenuated thermal component was included in the fits,
see text for details} 

\enddata
\end{deluxetable}

\section{Discussion} 

The HRI image of Mrk~231 shows emission across several tens of kpc, 
consistent with the extent of the host galaxy. The soft flux has 
two peaks of X-ray emission of similar 
strength in the 0.1 -- 2 keV band. These peaks are separated by 
$\sim 13''$ and the summed emission has 
$L(0.5-4.5\ {\rm keV}) \sim 10^{42} {\rm erg\ s}^{-1}$, possibly 
associated with starburst activity in the galaxy. 
The $60$ and $100 \mu m$ fluxes can be used to estimate the X-ray luminosity 
expected from starburst emission in this galaxy, following the 
prescription of David \etal\ (1992). Taking the IR fluxes from Smith, Lonsdale 
and Lonsdale (1998), an associated X-ray luminosity 
$L(0.5-4.5\ {\rm keV}) = 4.43^{+4.38}_{-1.86} \times 10^{42} {\rm erg\ s}^{-1}$ 
is estimated to originate from a starburst region of the observed IR 
luminosity. The 
PSPC flux yields $L(0.5-4.5\ {\rm keV})=1.43^+_-0.30  \times 10^{42} {\rm erg\ s}^{-1}$ 
within an extraction cell of radius $1.3'$. The fact the extended 
X-ray emission yields a lower luminosity than expected either supports 
the hypothesis that a significant fraction of the IR 
emission is powered by the active 
nucleus (Bryant \& Scoville  1996) or suggests that the X-ray emission from the 
starburst is attenuated by material which 
is more dusty than that observed 
in the David \etal\ (1992) sample of galaxies.

The observed {\it ASCA} spectrum shows a hard X-ray source exists 
in Mrk~231. Column 
measurements range from $\sim 2 \times 10^{22} - 10^{23} {\rm cm}^{-2}$ 
depending on whether the absorber is assumed to be neutral or ionized, 
and depending on assumptions about the underlying continuum form. 
In both cases an unattenuated component was included in the model 
to describe the extended soft X-ray emission. 
If the nuclear attenuation is as high as $N_H \sim 10^{23} {\rm cm}^{-2}$ 
then the intrinsic hard X-ray luminosity is $L(2-10\ {\rm keV}) \sim 10^{43} 
{\rm erg\ s}^{-1}$. Extrapolation of the {\it ASCA} fits to the 
50 -- 200 keV band yield an estimated $\gamma$-ray luminosity 
$L(50-200\ {\rm keV}) \sim 10^{43} {\rm erg\ s}^{-1}$, significantly lower than
the OSSE upper limit of $L(50-200\ {\rm keV}) < 2 \times 10^{45} {\rm erg\ s}^{-1}$ 
(Dermer \etal\ 1997; assuming our value for $H_{\circ}$).

The column density derived from fits which assume a neutral absorber 
are in approximate consistency with the 
value of $10^{22} {\rm cm}^{-2}$  estimated from the BAL absorbing systems 
(Rudy, Foltz and Stocke 1985) perhaps indicating an association 
with that gas. 
The presence of the 10$\mu m$ silicate absorption
feature and the relative levels of Na and Ca features (Rudy \etal\ 1985) 
along with the high polarization (Smith \etal\ 1995) 
show that there is a great deal of circumnuclear dust. 
Furthermore, Weymann \etal\ (1991) note that BAL clouds with high enough column 
density to produce the Fe{\sc ii} and Fe{\sc iii} absorption 
(as seen in Mrk~231) are consistent with the existence of dust 
within the clouds. 
C and O may be incorporated into molecules in such dusty material, 
and consequently depleted in the gas-phase compared to the 
levels assumed in models used here for the X-ray absorption, 
this may result in X-ray column measurements which underestimate the 
nuclear absorption by a factor of $\sim$ several.

\subsection{Is Mrk~231 X-ray quiet?}

It is interesting to investigate whether these new measurements of the nuclear 
absorption suggest Mrk~231 is intrinsically X-ray quiet, or whether 
the X-ray attenuation can account for its appearance as such. 
Green \etal\ (1995) found BAL QSOs to be X-ray weak using observed 
{\it ROSAT} PSPC  fluxes; i.e. 30 -- 100 times
less luminous in the 0.1 -- 2 keV band than would be expected by extrapolation
from their optical luminosities. The PSPC spectra did not allow Green
\etal\ (1995) to determine whether this X-ray weakness was due to 
absorption of the X-rays by a large column of material, 
or whether the sources were intrinsically weak. 
An {\it ASCA} observation of PHL 5200 (Mathur, Elvis and Singh 1995) had low 
signal-to-noise but indicated the presence of an absorbing column 
$N_H \sim 10^{23} {\rm cm^{-2}}$ attenuating the nuclear X-rays. 

Lawrence \etal\ (1997) compiled 
flux and line measurements across the broad-band 
spectrum for a sample of AGN with strong Fe{\sc ii} emission, and 
examined correlations between the various quantities. Those authors used 
the {\it ROSAT} PSPC data from Mrk~231 and derived indices 
$\alpha_{ix}= 1.97$ and $\alpha_{ox}=1.76$ between the $1\mu m$ and 2 keV 
flux, and between the 2500\AA\ and  2 keV flux, respectively.  
Mrk~231 was found to be a persistent outlier on their 
correlation plots, often due to the weakness in the X-ray band. 
Furthermore, Lawrence \etal\ (1997) found the 
 {\it ROSAT} PSPC spectrum to rule out the presence of a large 
column density ($N_H \simgreat 10^{21} {\rm cm^{-2}}$) 
for the X-ray absorber, producing a spectral paradox 
because a large column is required to bring the 
spectral-energy-distribution into line with other strong 
Fe{\sc ii} emitters. However, as previously discussed, the 
PSPC spectrum is complex and contains a significant contamination 
from a region of extended emission, which has a lower 
absorption than the nucleus. The PSPC is also 
insensitive to column densities $> 10^{22}  {\rm cm^{-2}}$. 
The {\it ASCA} data 
allow a more accurate determination of large 
absorbing columns and hence a more accurate determination of the  
intrinsic 2 keV flux from the nucleus. Combining this information with the 
{\it ROSAT} results we can subtract off the contamination and obtain 
a more accurate estimate of the nuclear flux at 2 keV. 
Thus the energy indices were recalculated using the {\it ASCA} 
measurement of the intrinsic nuclear flux at 2 keV;  
an absorption correction was applied and 
the estimated contribution to X-ray 
flux from the region of extended emission was subtracted. 

Smith \etal\ (1995) show a Faint Object Spectrograph (FOS) spectrum which has an observed flux 
$\sim 2 \times 10^{-27} {\rm erg\ cm^{-2}\ s^{-1}\ Hz^{-1}}$ at a rest-energy 
of 2500\AA. It is difficult to estimate the reddening-correction which 
should be applied to this 2500\AA\ flux; dust is thought to be an important 
constituent of the circumnuclear material, but there is no 2200\AA\ dust 
feature in the FOS spectrum (Smith \etal\ 1995). Laor and Draine (1993) 
show absorption cross-sections per H nucleus calculated assuming a range 
of differing dust grain types and size distributions. I 
assume a cross-section in the middle of their range of models, i.e. 
$\tau \sim 10^{-22} N_H^{-1} {\rm cm^2}$. If the lowest realistic 
value of X-ray 
absorption is taken, $N_H=2 \times 10^{22} {\rm cm^{2}}$ (Table~1), then a 
corrected 2500\AA\ 
flux $\sim 1.5 \times 10^{-26} {\rm erg\ cm^{-2}\ s^{-1}\ Hz^{-1}}$ 
is obtained. If $N_H \sim 10^{23} {\rm cm^{-2}}$ then 
the reddening-corrected flux is 
$\sim 4.6 \times 10^{-23} {\rm erg\ cm^{-2}\ s^{-1}\ Hz^{-1}}$  at 2500\AA. 

In Case-1, the {\it ASCA} data imply an intrinsic (absorption-corrected) 
nuclear flux 
of $1.7 \times 10^{-30} {\rm erg\ cm^{-2}\ s^{-1}\ Hz^{-1}}$ at 2 keV, 
assuming a power-law partially covered by neutral material 
(the first fit in Table~1), after 
subtracting the contamination from sources evident in the {\it ROSAT} 
data. Considering the range of possible values for absorption-corrected optical flux yields    
$1.51 < \alpha_{ox} < 2.85$. 
In  Case-2, the inferred 2 keV flux from the nucleus is 
$8.2 \times 10^{-31} {\rm erg\ cm^{-2}\ s^{-1}\ Hz^{-1}}$ at 2 keV, 
assuming the absorber is neutral material, and $\Gamma=1.5$ (fit 6 in Table~1). 
This case yields $1.64 < \alpha_{ox} < 2.97$. 
Assumption of an ionized absorber in Case-2, with $\Gamma=1.5$ 
(fit 9, Table~1), $N_H \sim 10^{23}$  yields a 2 keV  flux 
$\sim 9.3 \times 10^{-31} {\rm erg\ cm^{-2}\ s^{-1}\ Hz^{-1}}$ at 2 keV and 
$1.62 < \alpha_{ox} < 2.95$. Estimating the $1\mu m$ flux from 
Rigopoulou \etal\ (1996) yields $\alpha_{ix} \sim 1.7 - 1.8$ for 
Case-1 and Case-2, respectively.

These new indices do not bring Mrk~231  into the distribution of 
properties seen for other strong Fe{\sc ii} emitters 
(Lawrence \etal\ 1997). Furthermore, the lower limit in Case-1 lies just above 
the range $\alpha_{ox} \sim 1.1-1.5$ found for 
Seyfert 1 galaxies (Kriss \& Canizares 1985). 
However, all estimates of $\alpha_{ox}$ are consistent with the range 
$\alpha_{ox} \sim 1.3-1.9$ observed for the radio-quiet quasars, 
ie. the higher luminosity sources studied by  
Kriss \& Canizares (1985). 
As a Seyfert galaxy it would be 
considered to be at the X-ray-weak end of the distribution of indices for that class, but 
amongst QSOs it would not be considered X-ray-weak at all. 
The nominal ranges for $\alpha_{ox}$ combined 
with the absence of narrow emission lines in the optical bandpass, 
the presence of BAL systems and the high 
bolometric luminosity of Mrk~231 suggests the source is closer in nature to 
a QSO than a Seyfert galaxy. 

\section{Summary}

An HRI image of Mrk~231 shows extended emission in the 0.1 -- 2 keV band, 
of size-scale $\sim 30$ kpc in diameter, cospatial with the host galaxy. 
The nucleus itself appears faint in the soft X-ray band, probably due to 
absorption. The extended emission has a luminosity 
$L(0.1-2\ {\rm keV}) \sim 10^{42} {\rm erg\ s}^{-1}$ and is probably 
associated with starburst activity in the galaxy. The relative levels 
of IR and X-ray emission suggest some of the IR emission has an origin 
other than starbursts, or that the starburst region has a higher dust 
content than observed in other starburst galaxies. 

An {\it ASCA} observation revealed a hard X-ray source in Mrk~231. The 
spectrum indicates the X-ray column density 
is $N_H \sim 2 \times 10^{22} - 10^{23} {\rm cm}^{-2}$, attenuating the 
nuclear X-rays.
Consideration of the appropriate absorption-corrections to 
the optical and X-ray data infer an underlying 
spectral-energy-distribution for Mrk~231 which is more like that of a 
QSO than a Seyfert~1 galaxy.

\section{Acknowledgements}
I am grateful to \asca\ team for their operation of the satellite 
and to Ian George and Richard Mushotzky  for useful comments. 
Thanks to Keith Mason for confirmation of the optical positions 
of some AGN in the {\it ROSAT} field-of-view, and to Hagai Netzer 
for use of tables based upon ION. This research has 
made use of the NASA/IPAC Extragalactic database,
which is operated by the Jet Propulsion Laboratory, Caltech, under
contract with NASA; of the Simbad database, 
operated at CDS, Strasbourg, France; and data obtained through the 
HEASARC on-line service, provided by NASA/GSFC. I acknowledge the 
financial support of the Universities Space Research Association, 
University of Maryland, Baltimore County and LTSA grant NAG5-7385.

\newpage

\clearpage
\pagestyle{empty}
\setcounter{figure}{0}
\begin{figure}
\plotone{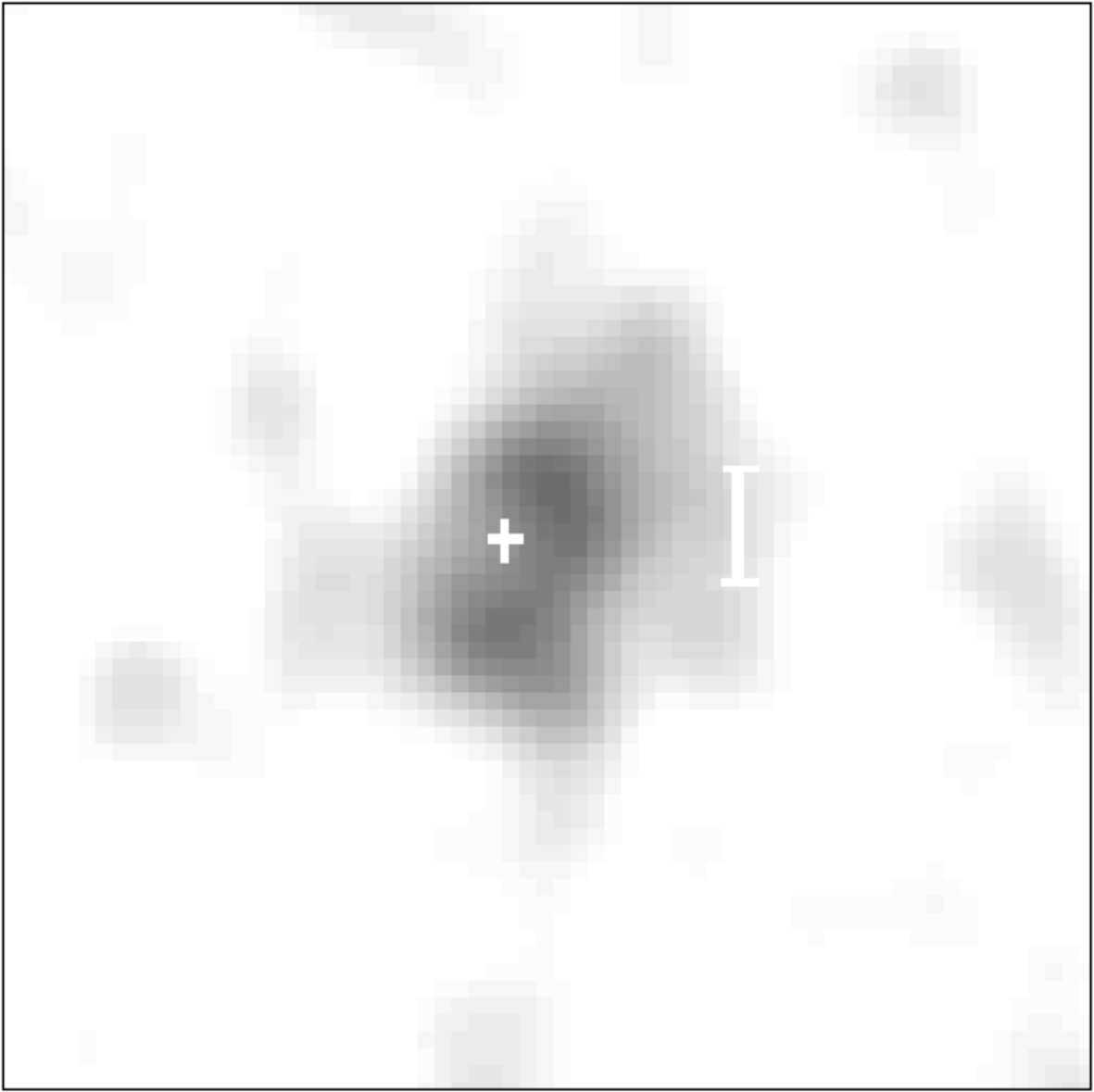}
\caption{
\label{fig:1}
The {\it ROSAT} HRI image of Mrk~231, in $1''$ pixels smoothed with a gaussian 
function with $\sigma=1.5''$. The orientation is such that 
North is at the top of the image, and East is to the left. 
The position of the radio nucleus is 
marked with a white cross, the bar represents $10''$ which is 
equal to $\sim 10$ kpc at the redshift of the source. }
\end{figure}
\clearpage

\clearpage
\pagestyle{empty}
\setcounter{figure}{1}
\begin{figure}
\plotone{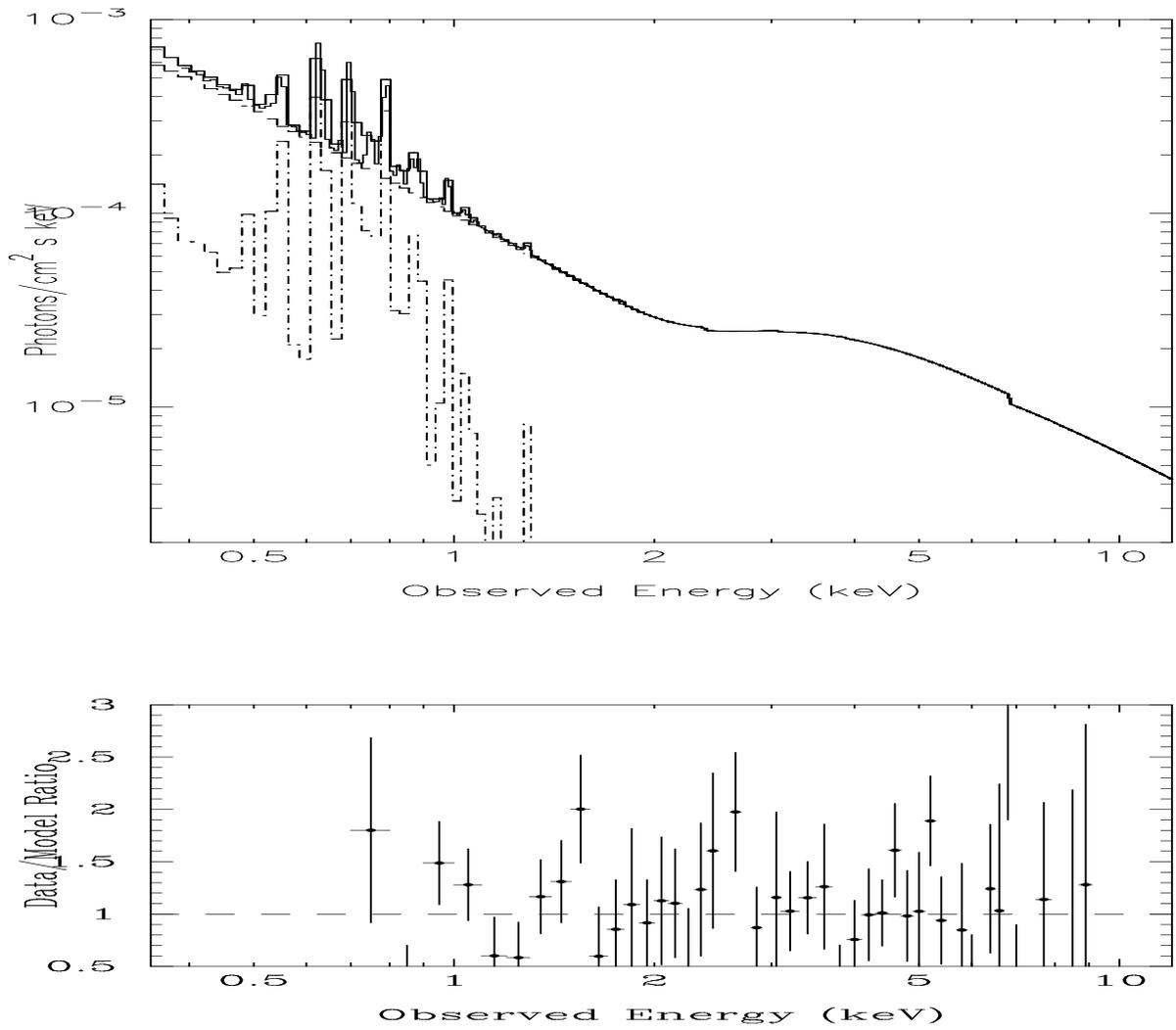}
\caption{
\label{fig:2}
{\it Top Panel}: The unfolded spectrum, from a fit to a power-law 
attenuated by an absorber partially covering the source (fit 1, 
see Table~1). 
{\it Bottom Panel}: 
The data/model ratio for the {\it ASCA} instruments, versus the
power-law model with index left free and a neutral absorber 
partially covering the source 
}
\end{figure}

\end{document}